# Spectral characteristics of the Fabry-Perot interferometer transmission upon illumination by an arbitrary light beam

A.Ya. Bekshaev, V.M. Grimblatov, O.N. Okunisnikov, R.A. Petrenko, V.N. Koverznev

Byelorussian SSR Academy of Science
Editorial board of the Zhurnal Prikladnoi Spektroskopii
Minsk 1982.

It is known that spectral properties of the Fabry – Perot interferometer (FPI) depend on the spatial characteristics of the incident radiation. This paper proposes a general method for taking this dependence into account, based on the decomposition of the incident field into the angular spectrum of plane waves, which are the FPI eigenfunctions. In the scalar approximation, the angular-frequency transfer function and the point-spread function (Green function) of the FPI are calculated, which enables to relate the spatial coherence functions of the incident and transmitted radiation and to derive an expression for the output intensity $I(\mathbf{r}, k)$ and for the integral transmittance $P(k) = \int I(\mathbf{r}, k)d\mathbf{r}$ depending on the wavenumber of the incident monochromatic radiation $k$. In the resulting expressions, $I(\mathbf{r}, k)$ is completely determined by the Fourier transform of the input coherence function, and $P(k)$ is determined by the angular power spectrum (APS) of the incident radiation. For beams of small divergence and high-quality FPIs, simplified formulas have been obtained that allow, in the first approximation, for the dependence of the FPI parameters on the angle of incidence and on the radiation wavelength.

Analysis of the expressions obtained makes it possible to clarify an exact meaning of the well-known regularities in the distortions of the FPI transmission band associated with the incident beam deviation from the plane wave (shift to the shortwave region, broadening, reduction of the maximum transmittance and the appearance of asymmetry). As examples, calculations of the integral transmittance for three types of APS, corresponding to Gaussian and conical beams, and the output intensity distribution $I(\mathbf{r}, k)$ for the input beam with a Gaussian APS are performed. Results are consistent with other theoretical and experimental work.

This paper in Russian was submitted to Journal of Applied Spectroscopy (Zhurnal Prikladnoi Spektroskopii, Minsk, USSR) in 1982, and had been accepted for publication. Because of the large volume, the journal published only the paper's Abstract [Zhurnal Prikladnoi Spektroskopii. – 1983. – V. 38, No. 5. – P. 866-867] while the text itself, according to the Editorial Board's decision 15 June, 1982, had been deposited in the VINITI database, Moscow, USSR (Деп. в ВИНИТИ. № 4251-82Деп.). Later discussions have revealed that some aspects considered in this paper can be of interest even now; especially, its approach can be applied to the study of spatial transformations performed by the FPI to structured input beams, including the singular and vortex beams. Therefore, its English translation is presented here.

**1. The Fabry – Perot interferometer** (FPI) is widely used as an efficient optical device for the spectral selection and transformation [1–3]. Its most important characteristic is the dependence of the FPI transmittance on the wavelength of the incident monochromatic light – the shape of the transmission band (transmission contour). As is known [2,3], the transmission contour depends on the spatial intensity distribution in the incident beam, and the usual FPI characteristics (e.g., it nominal bandwidth) implicitly imply that it "works" with a plane wave incident normally to its mirrors. At the same time, many of the FPI applications require that the spatial inhomogeneity of the incident beam be taken into account, e.g., when investigating the spectral composition of narrow laser beams, or when the FPI is employed as an interference filter in various opto-electronic systems. This had stimulated a series of works where the behavior of PFIs and interference filters was analyzed in different light beams with spatial inhomogeneity of the amplitude and phase distributions [4–9].

However, the accurate numerical calculations performed with account for specific structure of the multilayer filters [4–6] as well as approximate analytical approaches based on the standard FPI model [7–9] were applicable only to certain special forms of the incident beams. In this work we describe a rather simple analytical method for calculation of the FPI characteristics based on the incident beam representation via the angular spectrum of plane waves [10]. Immediately, the method is developed for the simplest FPI model including 2 thin mirrors characterized by their transmission and reflection coefficients but, with some evident complications stipulated by the properties of multi-layer dielectric mirrors, it can be generalized to the multilayer interference filters of light.

**2. Let us consider the FPI** as a linear optical system whose axis coincides with the interferometer axis; the input plane is situated immediately before the input mirror and the output plane – just behind the output one. In the scalar approximation (neglecting the light polarization effects), the following linear relation between the input $f(\mathbf{r}')$ and the output $g(\mathbf{r})$ complex amplitudes of the monochromatic optical field takes place:

$$g(\mathbf{r}) = \int f(\mathbf{r}') G(\mathbf{r},\mathbf{r}') d\mathbf{r}' \quad (1)$$

where $\mathbf{r}'$ and $\mathbf{r}$ are the 2D radius-vectors of points in the input and output planes, correspondingly; $G(\mathbf{r},\mathbf{r}')$ is the point-spread function [11], or the Green function (GF) of the optical system, and the integration over the whole input plane is implied in Eq. (1) (here and further, all such integrals are supposed to converge). In this work, we deal with a monochromatic radiation of frequency $\omega$, and the "true" time-dependent input and output fields are $\mathrm{Re}[f(\mathbf{r}')\exp(i\omega t)]$ and $\mathrm{Re}[g(\mathbf{r})\exp(i\omega t)]$.

Now we introduce the functions of spatial coherence $F(\mathbf{r}_1',\mathbf{r}_2') = \langle f(\mathbf{r}_1') f^*(\mathbf{r}_2') \rangle$ and $\Gamma(\mathbf{r}_1,\mathbf{r}_2) = \langle g(\mathbf{r}_1) g^*(\mathbf{r}_2) \rangle$ for the input and output fields (angular parentheses denote averaging). Then, Eq. (1) leads to the relation

$$\Gamma(\mathbf{r}_1,\mathbf{r}_2) = \int F(\mathbf{r}_1',\mathbf{r}_2') G(\mathbf{r}_1,\mathbf{r}_1') G^*(\mathbf{r}_2,\mathbf{r}_2') d\mathbf{r}_1' d\mathbf{r}_2' \quad (2)$$

which is an analog of the Hopkins formulas [12,13] written in the form suitable for our purpose. In the majority of experiments just the light intensity is interesting, so in what follows we will employ the consequences of formula (2):

$$I(\mathbf{r}) = \Gamma(\mathbf{r},\mathbf{r}) = \int F(\mathbf{r}_1',\mathbf{r}_2') G(\mathbf{r},\mathbf{r}_1') G^*(\mathbf{r},\mathbf{r}_2') d\mathbf{r}_1' d\mathbf{r}_2' , \quad (3)$$

$$P = \int I(\mathbf{r}) d\mathbf{r} = \int F(\mathbf{r}_1',\mathbf{r}_2') \left[ \int G(\mathbf{r},\mathbf{r}_1') G^*(\mathbf{r},\mathbf{r}_2') d\mathbf{r} \right] d\mathbf{r}_1' d\mathbf{r}_2' \quad (4)$$

where $I(\mathbf{r})$ is the intensity and $P$ is the total power of the transmitted radiation.

For the GF calculation we employ the fact that an incident plane wave with the input amplitude

$$f(\mathbf{r}') = \exp(i\mathbf{q}\mathbf{r}'), \quad (5)$$

after passing the FPI, produces the output plane wave with the same propagation direction,

$$g(\mathbf{r}) = H(\mathbf{q}) \exp(i\mathbf{q}\mathbf{r}) \quad (6)$$

where $H(\mathbf{q})$ is the transfer function, or "spatial-frequency response function" [10] of the FPI. Its expression is well known [1–3]:

$$H(\mathbf{q}) = \frac{T \exp\left(iL\sqrt{k^2 - q^2}\right)}{1 - R \exp\left[2i\Delta(\mathbf{q},k)\right]}, \tag{7}$$

$$\Delta(\mathbf{q},k) = L\sqrt{k^2 - q^2} + \frac{1}{2}\beta(\mathbf{q},k)$$

where $L$ is the distance between the mirrors of the FPI, $k$ is the radiation wavenumber in the medium between the mirrors, $T$ is the resulting amplitude transmission of the mirrors and the medium between them, $R$ and $\beta$ are the modulus and phase of the product $\rho_1\rho_2\tau^2$ in which $\rho_1$ and $\rho_2$ are the amplitude reflection coefficients of the mirrors 1 and 2, and $\tau$ is the amplitude transmission coefficient of the medium. Quantities $T$ and $R$ are, generally, complex and may depend on $\mathbf{q}$ and $k$.

Function $H(\mathbf{q})$ is an analogue to the frequency response of a linear electric circuit with constant parameters, just like the GF (point spread function) is an analogue of the impulse response of this circuit [10]. Therefore, the GF is a Fourier transform of (7),

$$G(\mathbf{r},\mathbf{r}') = \int \frac{T \exp\left(i\mathbf{q}(\mathbf{r}-\mathbf{r}') + iL\sqrt{k^2 - q^2}\right)}{1 - R \exp\left[2i\Delta(\mathbf{q},k)\right]} \frac{d\mathbf{q}}{(2\pi)^2}. \tag{8}$$

Substituting this into (3), one finds

$$I(\mathbf{r},k) = \int \frac{T^2 \Phi(\mathbf{q}_1,\mathbf{q}_2) \exp\left[i(\mathbf{q}_1 - \mathbf{q}_2)\mathbf{r} + iL\left(\sqrt{k^2 - q_1^2} - \sqrt{k^2 - q_2^2}\right)\right]}{\{1 - R\exp[2i\Delta(\mathbf{q}_1,k)]\}\{1 - R\exp[-2i\Delta(\mathbf{q}_2,k)]\}} \frac{d\mathbf{q}_1 d\mathbf{q}_2}{(2\pi)^4} \tag{9}$$

where $\Phi(\mathbf{q}_1,\mathbf{q}_2)$ is the double Fourier transform of the input function of coherence

$$\Phi(\mathbf{q}_1,\mathbf{q}_2) = \int F(\mathbf{r}_1',\mathbf{r}_2') \exp\left[-i(\mathbf{q}_1\mathbf{r}_1' - \mathbf{q}_2\mathbf{r}_2')\right] d\mathbf{r}_1' d\mathbf{r}_2'. \tag{10}$$

Integration of (9) over the whole output plane, with account for (4), gives

$$P(k) = \int \frac{T^2 \Phi(\mathbf{q},\mathbf{q})}{(1-R)^2 + 4R\sin^2\left[\Delta(\mathbf{q},k)\right]} \frac{d\mathbf{q}}{(2\pi)^2}. \tag{11}$$

Note that the function of coherence is positive semi-definite [13]. Since the Fourier transform, as a unitary transformation in the Hilbert space, does not change the function definiteness, $\Phi(\mathbf{q}_1,\mathbf{q}_2)$ is also positive semi-definite. Accordingly,

$$\Phi(\mathbf{q},\mathbf{q}) \geq 0. \tag{12}$$

If at the FPI input the incident field is coherent, $F(\mathbf{r}_1',\mathbf{r}_2')$ is factorizable, and then

$$\Phi(\mathbf{q}_1,\mathbf{q}_2) = A(\mathbf{q}_1) A^*(\mathbf{q}_2) \tag{13}$$

where $A(\mathbf{q})$ is the angular spectrum of the input field $f(\mathbf{r}')$. From now on, we will call the quantity $\Phi(\mathbf{q},\mathbf{q})$ "angular power spectrum" (APS). In case of statistically uniform fields this quantity actually coincides with the APS discussed in [10].

In order to derive the expression for the FPI transmission, we divide the expression (11) by the incident power, and represent the result in the form

$$\theta(k) = \int \frac{T^2}{(1-R)^2} \frac{B(\mathbf{q}) d\mathbf{q}}{1 + \frac{4R}{(1-R)^2}\sin^2 \Delta(\mathbf{q},k)} \tag{14}$$

where $B(\mathbf{q})$ is the "normalized" APS

$$B(\mathbf{q}) = \frac{\Phi(\mathbf{q},\mathbf{q})}{(2\pi)^2 P_0} \qquad (15)$$

with $P_0$ being the incident beam power, whence

$$\int B(\mathbf{q}) d\mathbf{q} = 1. \qquad (16)$$

According to (14), the only incident beam characteristic that determines the FPI transmittance is its APS. Recognition of this fact enables to develop a systematic approach to the problems of analyzing the FPI transmission in real beams. For example, it is known that the APS does not change upon the beam transmission in a free space [10]. That is why the FPI transmission band pattern does not depend on the beam cross section in which it is positioned. If an interference filter operates in a non-parallel light beam, it can be equally efficient in any beam cross section (practically, however, it is desirable to place the filter in the region of minimal cross section of the beam to avoid the influence of the filter non-uniformity).

Within the frame of the scalar approximation, Eqs. (9) and (14) supply exhaustive description of the FPI transmission for arbitrary incident beam. In particular, in case of the incident plane wave $f(\mathbf{r}') = \exp(i\mathbf{pr}')$ for which, as can be easily shown,

$$B(\mathbf{q}) = \delta(\mathbf{p}-\mathbf{q})$$

where $\delta(\mathbf{p}-\mathbf{q})$ is the 2D $\delta$-function, it follows from (14),

$$\theta(k) = \theta_0(k) = \frac{T^2}{(1-R)^2} \frac{1}{1 + \frac{4R}{(1-R)^2}\sin^2\Delta(\mathbf{p},k)}, \qquad (17)$$

that is, the usual Airy distribution [12] (if $R$, $T$ and $\beta$ depend on $\mathbf{q}$, in (17) these are taken at $\mathbf{q} = \mathbf{p}$). The maximum value of (17) is reached at $k = k_1(\mathbf{p})$ such that $\Delta(\mathbf{p},k_1) = n\pi$ (integer $n$) and equals to

$$\theta_0(k_1) = \frac{T^2(\mathbf{p},k_1)}{\left[1-R(\mathbf{p},k_1)\right]^2}. \qquad (18)$$

Here, possible dependence of $R$ and $T$ on the angle of incidence and on the wavelength is reflected explicitly. Further, the general theorems of the integral calculus with the help of equalities (12), (15) and (16), dictate that for arbitrary $k$

$$\theta(k) \le \int \frac{T^2}{(1-R)^2} B(\mathbf{q}) d\mathbf{q} = \left(\frac{T}{1-R}\right)^2_a \qquad (19)$$

where the subscript "$a$" denotes that $\left[T/(1-R)\right]^2$ is taken at the certain value of $\mathbf{q}$ that belongs to the area of the $\mathbf{q}$-plane producing the main contribution to the integral. If $T$ and $R$ can be considered independent of $\mathbf{q}$, confronting of (18) and (19) yields

$$\theta(k) \le \theta_0(k_1) \qquad (20)$$

– that is, every broadening of the APS with respect to the delta-function APS of a plane wave) reduces the transmission peak. In general case, for variable $R$ and $T$, this conclusion looses its absolute character: in principle, one can find such a plane wave and a beam with such an APS that inequality (20) reverses. However, in practice, the $R$ and $T$ dependence of the incidence angle and frequency is employed with the opposite purpose: namely, to increase the transmission maximum for the chosen plane wave. Obviously, in such situations the inequality (20) will only be reinforced.

For practical calculations by means of Eqs. (9) and (14), one should know the dependences $R(\mathbf{q},k)$, $T(\mathbf{q},k)$ and $\beta(\mathbf{q},k)$. Within the frame of the proposed method, this dependence is introduced phenomenologically for a concrete structure of the mirrors and of the medium between them. The problem can be simplified if there is a limited region of the **q**-plane near the optical axis,

$$|\mathbf{q}| = q \ll k, \tag{21}$$

that produces the main contribution to the integrals (9) and (14), and if the transmission peak is narrow enough, i.e. all the values of $k$ important for the calculation obey the condition

$$|\kappa| \ll k_0, \quad \kappa = k - k_0, \tag{22}$$

where $k_0 = k_1(0)$ is the wavenumber at which the transmission is maximum for normally incident plane waves. Under conditions (21) and (22), the functions $R(\mathbf{q},k)$, $T(\mathbf{q},k)$ and $\beta(\mathbf{q},k)$ should be determined only within small regions of the **q**-plane and $k$-line where appropriate approximations are applicable [3]. For example, if the mirrors are metallic and $(q/k_0) < 0.1$, $(\kappa/k_0) < 0.01$, the quantities $R$, $T$ and $\beta$ are constant with high accuracy [3]. If the mirrors are dielectric multilayer structures, at least partial allowance for the angular and frequency dependence of their parameters is necessary. According to [3], the main influence is coupled with the spectral and angular dispersion of the phase shift $\beta$, whereas the variations of $R$ and $T$ in the regions defined by conditions (21), (22) are small because the mirrors are usually fabricated so that near $q = 0$, $k = k_0$ the functions $R$ and $T$ possess extrema. Therefore, in further consideration we take into account the variations of $\beta$ within the first non-vanishing degrees of the power expansion in $\kappa$ and $q$ while $R$ and $T$ will be supposed constant.

Then,

$$\Delta(\mathbf{q},k) \approx k_0 L + \frac{1}{2}\beta(0,k_0) + k_0 L\left(\frac{\kappa}{k_0} - \frac{q^2}{2k_0^2}\right) + \frac{1}{2}\frac{\partial\beta}{\partial k}\kappa + \frac{1}{2}\frac{\partial\beta}{\partial(q^2)}q^2$$

$$= n\pi + k_0\left(L + \frac{1}{2}\frac{\partial\beta}{\partial k}\right)\frac{\kappa}{k_0} - k_0\left[L - k_0\frac{\partial\beta}{\partial(q^2)}\right]\frac{q^2}{2k_0^2} \tag{23}$$

because, in the scalar approximation, the angular dependence of any quantities cannot contain terms of the first degree in **q**. The derivatives of $\beta$ are taken in the point $\mathbf{q} = 0$, $k = k_0$. Now, supposing the second and third summands of (23) to be much less than unity, we transform Eq. (14) to the form

$$\theta(k) = \frac{T^2}{(1-R)^2}\bar{\theta}(\sigma), \quad \bar{\theta}(\sigma) = \int\frac{B(\mathbf{u})d\mathbf{u}}{1+(\sigma-u^2)^2} \tag{24}$$

where

$$\sigma = \frac{\kappa}{\gamma k_0}, \quad \mathbf{u} = \frac{\mathbf{q}}{\sqrt{2\gamma}\,k_0}\nu, \quad \nu = \left(\frac{L - k_0\dfrac{\partial\beta}{\partial(q^2)}}{L + \dfrac{1}{2}\dfrac{\partial\beta}{\partial k}}\right)^{1/2}. \tag{25}$$

Here

$$\gamma = \frac{1-R}{2k_0\sqrt{R}}\left(L + \frac{1}{2}\frac{\partial\beta}{\partial k}\right)^{-1} \tag{26}$$

is the relative half-width of the transmission band for the normally incident plane waves. In Eq. (24), the APS is normalized in a new manner (cf. (15)):

$$B(\mathbf{u}) = \frac{\Phi(\mathbf{u},\mathbf{u})}{\int\Phi(\mathbf{u},\mathbf{u})d\mathbf{u}} = B(\mathbf{q})\frac{2\gamma k_0^2}{\nu^2}. \tag{27}$$

Note that expression (20) noticeably differs from zero only for small enough values of $\sigma$ and $u^2$ that do not exceed the unity by the order of magnitude. Since, in agreement with (26), for $(1-R) \ll 1$

$$\gamma \ll \left(k_0 L + \frac{1}{2} k_0 \frac{\partial \beta}{\partial k}\right)^{-1},$$

then, for small $\sigma$

$$k_0 \left(L + \frac{1}{2} \frac{\partial \beta}{\partial k}\right) \frac{\kappa}{k_0} = \gamma k_0 \left(L + \frac{1}{2} \frac{\partial \beta}{\partial k}\right) \sigma \ll 1.$$

Similar reasoning substantiates the smallness of the last term in (23) thus justifying the replacement of the sine function in (14) by its argument.

The simple expression (24) enables easily finding of the important characteristics of the transmission band. In particular, during the integration over $\sigma$ across the near-peak area, one may neglect the contribution of high $\sigma$ values supposing the integration limits infinite. Thus we determine the "peak area"

$$S = \int_{-\infty}^{\infty} \bar{\theta}(\sigma) d\sigma = \int_{-\infty}^{\infty} \frac{d\sigma}{1 + (\sigma - u^2)^2} = \pi. \tag{28}$$

It has been shown above that for any beam different from the plane wave, the peak maximum decreases. For the constant peak area (28) that means that its half-width grows, which leads, e.g., to the deterioration of the interference filter resolving power.

The peak position on the wavenumber axis can be, to a certain degree, characterized by the "center of gravity"

$$\bar{\sigma} = \frac{1}{S} \int_{-\infty}^{\infty} \sigma \bar{\theta}(\sigma) d\sigma.$$

According to (24)

$$\bar{\sigma} = \frac{1}{\pi} \iint \frac{\sigma B(\mathbf{u}) d\mathbf{u}}{1 + (\sigma - u^2)^2} d\sigma = \int u^2 B(\mathbf{u}) d\mathbf{u} = \overline{u^2}, \tag{29}$$

or, in the wavenumber units,

$$\overline{\kappa} = \gamma k_0 \overline{u^2} = \frac{\overline{q^2} v^2}{2 k_0},$$

where $\overline{q^2}$ is the mean square of the APS spatial frequency. The result (29) means that any broadening of the incident angular spectrum always induces the transmission band shift to the higher-wavenumber (short-wavelength) direction.

Now we consider the FPI band displacement when the incident beam, as a whole, is inclined with respect to the FPI axis (oblique incidence) [4–6]. This corresponds to the APS shift by a certain vector **p**. In the **u**-plane this shift is described by the vector

$$\mathbf{c} = \frac{v}{k_0 \sqrt{2\gamma}} \mathbf{p} \tag{29a}$$

and the APS transformation can be expressed as $B(\mathbf{u}) \to B(\mathbf{u} - \mathbf{c})$. Then

$$\bar{\sigma} = \int u^2 B(\mathbf{u} - \mathbf{c}) d\mathbf{u} = \int (\mathbf{v} + \mathbf{c})^2 B(\mathbf{v}) d\mathbf{v} = |\mathbf{c}|^2 + 2\mathbf{c} \int \mathbf{u} B(\mathbf{u}) d\mathbf{u} + \overline{u_0^2} \tag{30}$$

where $\overline{u_0^2}$ is the "center of gravity" displacement that would occur at $\mathbf{c} = 0$, i.e. if the incident beam is not inclined. Expression (30) acquires the especially simple form when $B(\mathbf{u})$ is an even function, e.g., if $B(\mathbf{u})$ possesses a circular symmetry. In this case, the peak displacements due to the symmetric APS broadening and due to the incident beam inclination as a whole are additive.

Substitution of (23) into (9) yields a useful representation for the output intensity. After the replacements similar to (25), and keeping the terms with orders not higher than $\sigma^2$ and $u^4$, a rather cumbersome expression is obtained that can be simplified if one assumes $v = 1$. In this case

$$I(\mathbf{r},\sigma) = \left(\frac{T}{1-R}\right)^2 \frac{\left(2k_0^2\gamma\right)^2}{(2\pi)^4} \int \frac{\Phi(\mathbf{u}_1,\mathbf{u}_2)\exp\left[i\vec{\xi}(\mathbf{u}_1-\mathbf{u}_2)\right]d\mathbf{u}_1 d\mathbf{u}_2}{1-ia(u_2^2-u_1^2)-d(u_2^2-u_1^2)^2+(\sigma-u_1^2)(\sigma-u_2^2)} \tag{31}$$

where

$$\vec{\xi} = \sqrt{2\gamma}\,k_0\mathbf{r}, \quad a = \frac{1+R}{2\sqrt{R}}, \quad d = \frac{(1-R)^2}{8R}.$$

The presented results agree with the experimental data as well as with the results of numerical analysis [4–6,8,9]. Possible refinements can be achieved after the more complete allowance for the variations of the mirrors' properties with the angle of incidence and the wavelength. Notably, the calculation of the FPI transmission, based on the integrals (9) and (14), is performed in the unified manner rather than by numerical methods reported earlier. The main convenience of the obtained expressions is their compactness and physical transparency due to which the roles of different beam and FPI parameters are plainly exposed, and the expressions can be efficiently investigated by analytical methods. Although less accurate in comparison to the numerical approaches, the calculations via the formulas (24) and (31) are much simpler and enable the express estimates.

The main drawback of the described approach is its scalar character and abstracting from the light polarization, which makes it inaccurate if the APS components with the incidence angles 5° – 10° have to be considered. This limitation agrees with other ones assumed in this work, and it does not restrict the applicability of the results obtained. Expectably, the approach can be generalized to the vector situations where the complex amplitudes are represented by 2D vectors, GF – by 2×2 matrices, and the coherence matrices take the places of the coherence functions [12,13].

**3. In this Section**, the general expressions derived above are employed for the analysis of the FPI transmission band for several beam types characterized by the special forms of their APS.

**3.1.** Let us start with the <u>uniform APS</u>:

$$\Phi(\mathbf{q},\mathbf{q}) = \begin{cases} \text{const}, & |\mathbf{q}| \leq q_0; \\ 0, & |\mathbf{q}| > q_0. \end{cases} \tag{31a}$$

Such form of the APS is typical, e.g., for the radiation obtained after the plane wave passes a circular lens of the limited transverse size, situated perpendicularly to the incident beam axis; here

$$q_0 = \frac{ka_0}{|f_0|} = k\psi \tag{31b}$$

with $a_0$ being the lens radius, $f_0$ its focal distance and $\psi$ is the convergence (divergence) angle of the focused (defocused) incident beam after its refraction at the FPI input face. In the dimensionless variables (25) this APS is characterized by

$$\alpha_p = \frac{q_0^2}{2\gamma k_0^2}v^2, \quad B(\mathbf{u}) = \begin{cases} (\pi\alpha_p)^{-1}, & u^2 \leq \alpha_p; \\ 0, & u^2 > \alpha_p. \end{cases} \tag{32}$$

Because in this case, the problem of the transmission band calculation is formally equivalent to the known problem of the FPI behavior in the converging beam [3,7], we just present the main results for further comparison with other cases. The integration in (24) with account for (32) gives

$$\bar{\theta}_p(\sigma) = \frac{1}{\alpha_p}\left[\arctan(\alpha_p - \sigma) + \arctan\sigma\right]. \tag{33}$$

The transmission maximum takes place at

$$\sigma = \sigma_{\max} = \frac{1}{2}\alpha_p \tag{34}$$

and equals to

$$\bar{\theta}_{p\max} = \frac{2}{\alpha_p}\arctan\frac{\alpha_p}{2}. \tag{35}$$

The transmission band center of gravity (29) is determined by

$$\overline{u^2} = \frac{1}{2}\alpha_p \tag{36}$$

and coincides with $\sigma_{\max}$ (35). The transmission contour is symmetric with respect to $\sigma_{\max}$, which enables to define its half-width $\gamma_1$ by means of the condition

$$\bar{\theta}_p(\sigma_{\max} \pm \gamma_1) = \frac{1}{2}\bar{\theta}_{p\max}, \quad \gamma_1 = \sqrt{1 + \frac{1}{4}\alpha_p^2}.$$

Approximately, these results were obtained in [3]. Fig. 1 illustrates them. Note that, in agreement with the general conclusions, when $\alpha_p$ grows, the transmission band moves to the higher $\sigma$ (shorter wavelengths), broadens and becomes essentially non-Lorentzian.

**3.2.** <u>Gaussian APS</u>. In the dimensionless variables

$$B(\mathbf{u}) = \frac{1}{\pi\alpha_G}\exp\left(-\frac{u^2}{\alpha_G}\right). \tag{37}$$

Such form of the APS is characteristic, e.g., for a Gaussian beam with the complex amplitude distribution in the input plane

$$f(\mathbf{r}') = \exp\left(-\frac{r'^2}{2b^2}\right) \tag{38}$$

where

$$b^2 = \frac{v^2}{2\alpha_G k_0^2 \gamma}. \tag{39}$$

By substituting (38) into (24) and employing the polar coordinates $u = |\mathbf{u}|$ and the polar angle $\mu$, $d\mathbf{u} = u\,du\,d\mu$, we arrive at

$$\bar{\theta}_G(\sigma) = \frac{1}{\pi\alpha_G}\int_0^\infty \frac{\exp(-u^2/\alpha_G)u\,du}{1+(\sigma-u^2)^2}\int_0^{2\pi}d\mu,$$

which via the replacement $u^2 = t$ leads to

$$\bar{\theta}_G(\sigma) = \frac{1}{\alpha_G}\int_0^\infty \frac{\exp(-t/\alpha_G)dt}{1+(t-\sigma)^2} = -\frac{1}{\alpha_G}\mathrm{Im}\left[e^Z E_1(Z)\right] \tag{40}$$

where $Z = \dfrac{i-\sigma}{\alpha_G}$, $E_1(Z)$ is the integral exponent [14]. Upon the condition $\alpha_G \to 0$, the expression (40) can be written as

$$\bar{\theta}_G(\sigma) \approx \frac{1}{1+\sigma^2}\left(1 + \frac{2\alpha_G\sigma}{1+\sigma^2}\right)$$

whence for the peak maximum position we obtain

$$\sigma^0_{max} \approx \alpha_G. \qquad (41)$$

At $\alpha_G \to \infty$ we employ the series expansion of the function $E_1(Z)$ [14] which gives

$$\bar{\theta}_G(\sigma) \approx \frac{1}{\alpha_G}\exp\left(-\frac{\sigma}{\alpha_G}\right)\cos\left(\frac{\pi - \arctan\sigma}{\alpha_G}\right). \qquad (42)$$

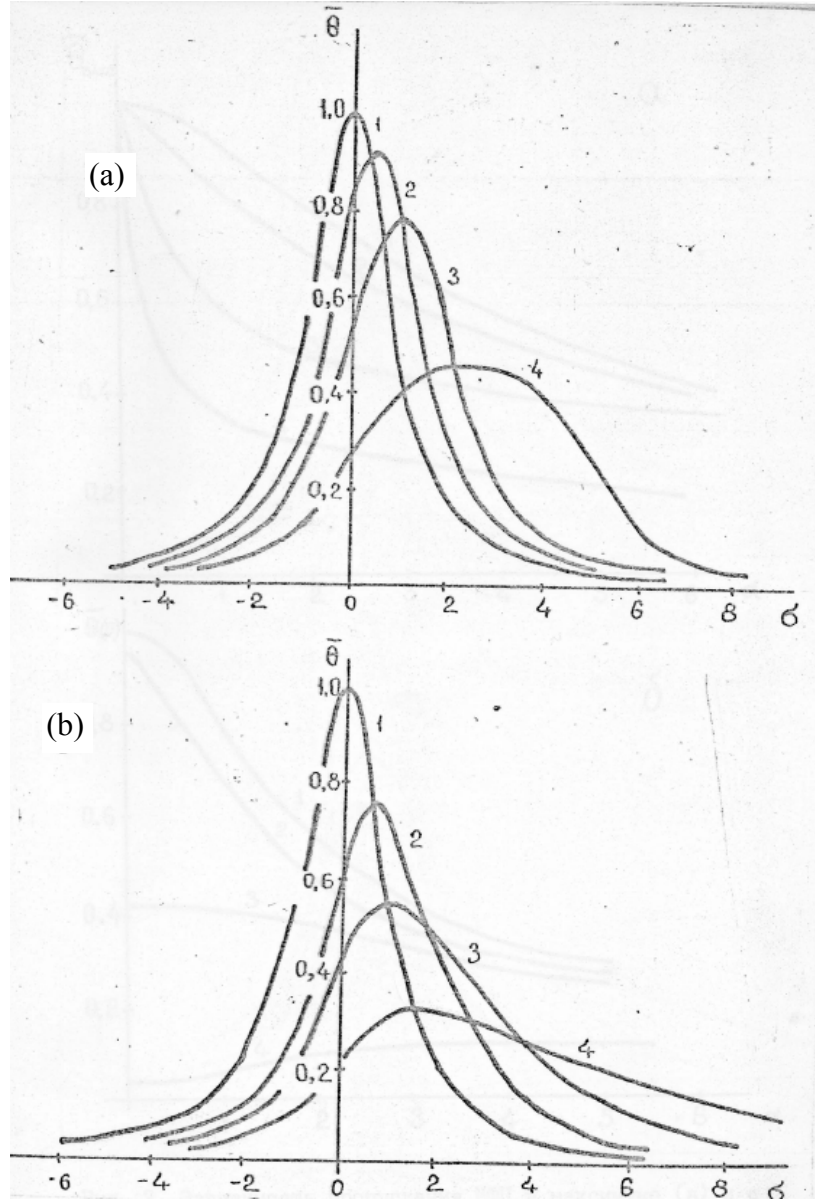

Fig. 1. Transmission band contours of the FPI illuminated by a beam with (a) uniform (31a) and (b) Gaussian (37) APS of different widths: (1) $\alpha_p, \alpha_G = 0$; (2) $\alpha_p, \alpha_G = 2$; (3) $\alpha_p, \alpha_G = 2$; (4) $\alpha_p, \alpha_G = 5$.

This function is sharply asymmetric, its graph steeply falls down when $\sigma$ decreases and smoothly decays when $\sigma$ grows. The position of the maximum at $\alpha_G \to \infty$ is determined by approximate relation

$$\sigma = \sigma^\infty_{max} = \frac{\alpha_G}{\pi}.$$

This behavior is similar to that described by (34) for the peak maximum position in case of the uniform APS. The center of gravity shift for arbitrary $\alpha_G$ equals to

$$\overline{u^2} = \alpha_G.$$

Obviously, at small $\alpha_G$, in the first approximation, the shift of the maximum (41) coincides with the center of gravity displacement but at larger $\alpha_G$, the center of gravity is shifted stronger. This additionally testifies for the transmission contour asymmetry at large $\alpha_G$. From comparing the results of the accurate calculations based on (40) (see Fig. 1b) with the uniform-APS case (Fig. 1a), one can remark that the peculiar feature of the Gaussian APS is the band contour asymmetry which grows with increasing $\alpha_G$.

The Gaussian beam size, at which the band contour distortion is noticeable, can be determined supposing $\alpha_G = 1$ in (39). For $k_0 = 10^5$ cm$^{-1}$ (He-Ne laser radiation) and $\nu = 1$

$$b \simeq b_0 = \frac{10^{-5}}{\sqrt{\gamma}} \text{ cm}.$$

For beams satisfying this condition, application of high-quality FPI may not be efficient because the bandwidth will be determined by the incident beam APS rather than by the FPI itself.

**3.3.** Uniform APS with a displacement:

$$B(\mathbf{u}) = \begin{cases} (\pi\alpha)^{-1}, & |\mathbf{u}-\mathbf{c}| \le \sqrt{\alpha}; \\ 0, & |\mathbf{u}-\mathbf{c}| > \sqrt{\alpha}. \end{cases} \qquad (43)$$

The simple example of such an APS is an oblique converging/diverging incident beam for which

$$c = |\mathbf{c}| = \frac{\nu}{\sqrt{2\gamma}} \sin\varphi$$

where $\varphi$ is the angle of incidence of the axial ray. Then, after the substitution of (43) into (24) and replacement $\mathbf{v} = \mathbf{u} - \mathbf{c}$ we have

$$\overline{\theta}_{pc}(\sigma) = \frac{1}{\pi\alpha} \int \frac{v dv}{1 + (v^2 + 2cv\cos\mu + c^2 - \sigma)^2} \qquad (44)$$

(here, again, the polar coordinates in the $\mathbf{v}$-plane are introduced, $v = |\mathbf{v}|$). The integral is calculated in polar coordinates over the circle centered at the origin and with radius $\sqrt{\alpha}$. At small $\alpha$, which corresponds to the small convergence angles, Eq. (44) can be represented via the power series in $\alpha$, and keeping the linear terms we find

$$\overline{\theta}_{pc}(\sigma) = \frac{1}{1+x^2} - \frac{\alpha}{(1+x^2)^2}\left(c^2 + x - \frac{4c^2 x^2}{1+x^2}\right) \qquad (45)$$

where $x = c^2 - \sigma$. According to (45), at $\alpha = 0$ (incident plane wave) the transmission contour does not change the shape compared to the case of normal incidence but is only shifted by $\sigma = c^2$. At $\alpha > 0$, the shape distortion is described by the additive term that behaves differently regarding the value of $\sigma$. The transmission maximum is still located at $\sigma = c^2$ and equals to

$$\overline{\theta}_{pc\max} = 1 - \alpha c^2 \qquad (46)$$

that is, at $\alpha > 0$ the transmission peak height reduces with growing convergence angle, and the speed of reducing, for small $\alpha$, is proportional to the square of the incidence angle. Corresponding dependences calculated accurately via Eq. (44) (Fig. 2a) illustrate this conclusion.

When the FPI serves as an interference filter, it is usually designed so that the transmission for the normally incident plane waves of a fixed prescribed wavelength be maximal. Due to the spatial inhomogeneity of the incident beam, the FPI transmittance for the prescribed wavelength decreases. This decrease is described by the quantity $\bar{\theta}(0)$. At small $\alpha$, Eq. (45) dictates

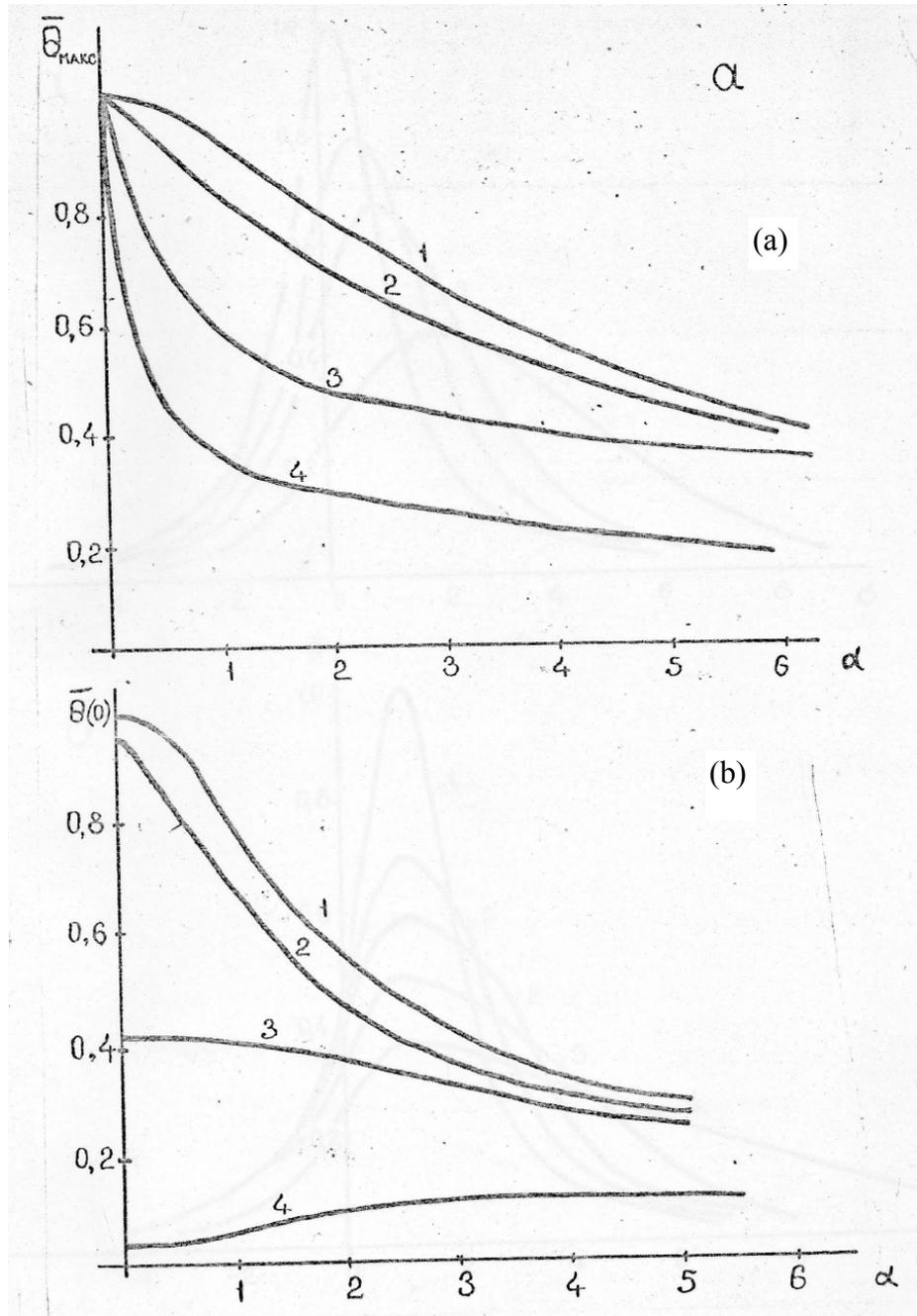

Fig. 2. FPI transmission (a) at the actual maximum $\bar{\theta}_{pc\,max}$ and (b) at the nominal band center $\bar{\theta}_{pc}(0)$ vs the uniform APS width for various incident beam inclinations:
(1) $c = 0$; (2) $c = 0.5$; (3) $c = 1.1$; (4) $c = 2.2$.

$$\bar{\theta}_{pc}(0) = \frac{1}{1+c^4} - \frac{2\alpha c^2 (1-c^4)}{(1+c^4)^3}, \qquad (47)$$

or, for small $c$,

$$\bar{\theta}_{pc}(0) \approx 1 - c^4 - 2\alpha c^2. \qquad (48)$$

Confrontation of Eqs. (46) and (48) shows that, at small incidence angles and small incident beam convergence, the FPI transmission for the nominal wavelength decreases twice as much as the transmission in the actual maximum does.

The exact dependence of $\bar{\theta}_{pc}(0)$ on the APS width $\alpha$ calculated via (44) is shown in Fig. 2b. Noteworthy, the $\bar{\theta}_{pc}(0)$ reduction with growing $\alpha$, observed for $c < 1$ (curves (1, 2), changes to the increase when $c > 1$ (curves 3, 4). This agrees with (47) and can be explained by the fact that when the incident conical beam is inclined, this may enlarge the number of rays approaching the FPI face at small angles (almost normally).

Expression (44) can also be simplified if $c \ll 1$. Keeping the second order in $c$ we find

$$\bar{\theta}_{pc}(\sigma) = \frac{1}{\alpha}\left[\arctan(\alpha - \sigma) + \arctan\sigma\right] - \frac{2c^2(\alpha - \sigma)}{\left[1 + (\alpha - \sigma)^2\right]^2}. \tag{49}$$

One can easily persuade that in the region of $\alpha$ and $c$ values where (47) and (49) are simultaneously valid, these equations lead to the coinciding results. From (49) it follows:

$$\sigma_{max} = \frac{\alpha}{2} + c^2 \frac{1 - \frac{1}{2}\alpha^2}{\left(1 + \frac{1}{4}\alpha^2\right)^2} \tag{50}$$

$$\bar{\theta}_{pc\,max} = \frac{2}{\alpha}\arctan\frac{\alpha}{2} - \frac{\alpha c^2}{\left(1 + \frac{1}{4}\alpha^2\right)^2} \tag{51}$$

$$\bar{\theta}_{pc}(\sigma) = \frac{1}{\alpha}\arctan\alpha - \frac{2c^2\alpha}{\left(1 + \alpha^2\right)^2} \tag{52}$$

As could be expected, when $c = 0$, Eqs. (49) – (51) reduce to Eqs. (33) – (35). The relations (45) and (49) supply explicit manifestations of the band asymmetry since their additional summands are non-symmetric with respect to the point of maximum and cause the broadening of the transmission contour to the direction of large $\sigma$. Not that, according to (50), when $\alpha^2 > 2$, the shift of the maximum for the oblique incident beam is smaller than for the normal one. Besides, at $\alpha \ll 1$ as well as at $c^2 \ll 1$, the shift of the maximum is smaller than the shift of the center of gravity: the latter, in agreement with (30) and (36) equals to

$$c^2 + \frac{1}{2}\alpha.$$

Numerical calculations performed along Eq. (44) confirm the above conclusions based on the approximate analysis (Fig. 3a, b). Like in the case of normal incidence (Fig. 2), with growing $\alpha$ the transmission band broadens and its maximum diminishes. However, the band contour changes asymmetrically: it acquires the articulate "tail" in the short-wavelength region but abruptly "drops" on the long-wavelength side. This is quite explainable from the geometric pattern: a significant fraction of plane waves composing the beam impinges the FPI at angles for which the transmission is maximum for the shorted waves. More interesting is the fact that with growing $c$, that is, with growing angle of incidence, the "blue" shift of the maximum caused by the APS broadening (finite value of $\alpha$) initially decreases, then disappears and ultimately inverses: the blue shift is replaced by the red one (Fig. 4). This occurs because at $\alpha > 0$ the incident beam angular spectrum contains plane waves which approach the FPI at angles lower than the incidence angle of the axial ray and for which the maximum transmission is displaced to the long-wave direction.

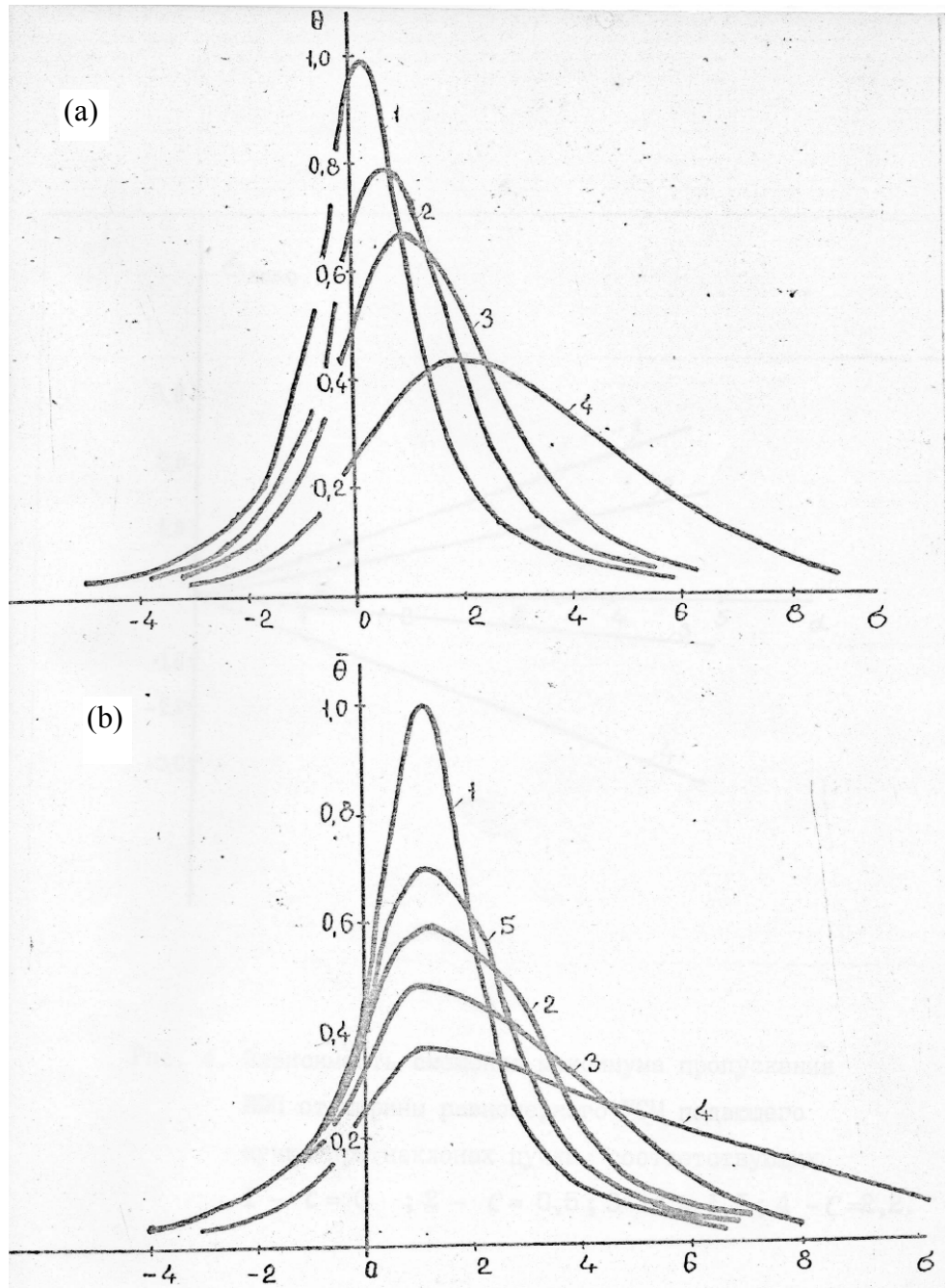

Fig. 3. Contours of the FPI transmission band for the incident beam with the uniform APS, when the incident beam axis inclination is characterized by (a) $c = 0.5$ and (b) $c = 1.1$. Different curves correspond to the beams with (1) $\alpha = 0$ (plane wave); (2) $\alpha = 1$; (3) $\alpha = 3$; (4) $\alpha = 5$; (5) $\alpha = 0.5$.

Note the essential growth in the band-shape sensitivity to the APS broadening with increasing $c$. A direct indication to this property is already present in expression (45) and (46). It takes place due to the same reason that causes the transmission band broadening for the normally incident beams. If the incident beam differs from the plane wave, the secondary beams obtained via the partial reflection from the FPI mirrors interfere with lowering contrast – no matter, because of the incomplete coherence or due to geometric discrepancies. In case when such a beam falls obliquely, the geometric discrepancies between the secondary beams substantially increase, and the speed of the band contour distortion increases too. One can readily imagine conditions at which the secondary beams practically do not interfere; then, the FPI transmission $\bar{\theta}$ does not depend on the wavelength and equals to the mean value of the Airy function [13] $(1-R)(1+R)^{-1}$.

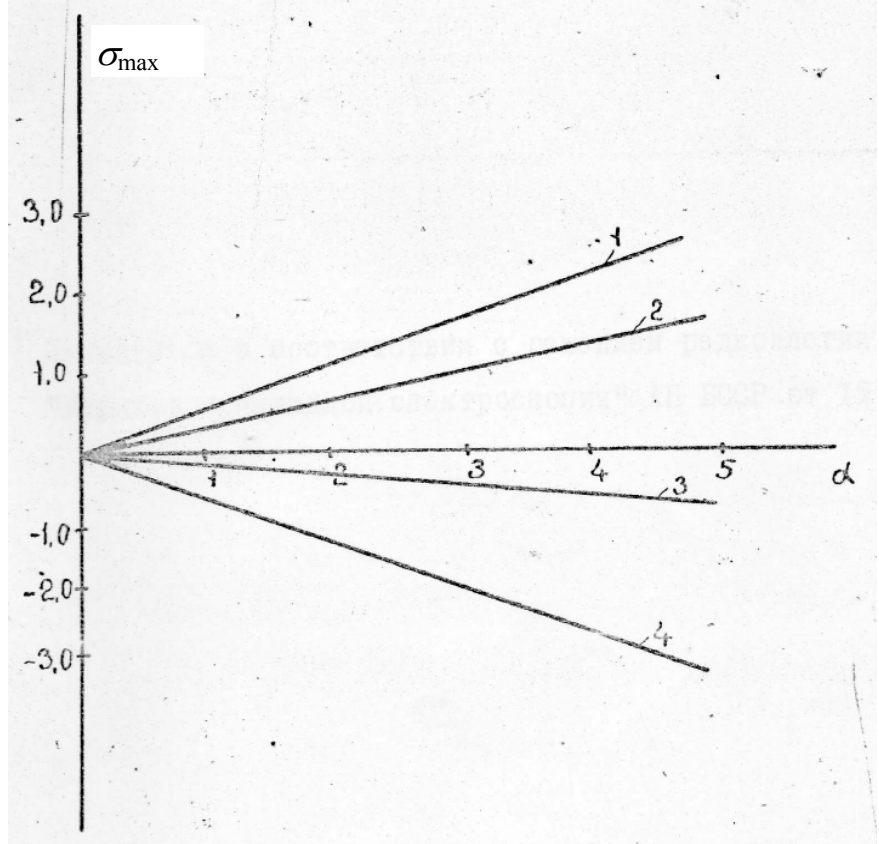

Fig. 4. Shift of the FPI transmission maximum as a function of the APS width $\alpha$ for incident beams with the uniform APS and different inclination angles:
(1) $c = 0$ (normal incidence); (2) $c = 0.5$; (3) $c = 1.1$; (4) $c = 2.2$.

**3.4.** So far, our attention was focused on the total output power (11) and, correspondingly, on the integral FPI transmission (14). However, the characteristic advantage of the method presented in the above sections is that it can be applied for the problems of spatial structure transformations induced by the FPI. Generally, this topic is outside the scope of the present paper but here we briefly consider an example illustrating the calculation of the light field spatial distribution at the FPI output. Let us take the input radiation in the Gaussian-beam form for which

$$\Phi(\mathbf{u}_1,\mathbf{u}_2) = 2\pi b^2 \exp\left(-\frac{u_1^2 + u_2^2}{2\alpha_G}\right).$$

After the substitution of this equality into (31), we obtain the general expression for $I(\mathbf{r},\sigma)$. In the limit $\alpha_G \to 0$ it gives

$$I(\mathbf{r},\sigma) = \frac{T^2}{(1-R)^2}\frac{1}{1+\sigma^2}\left[1+\frac{4\alpha_G \sigma}{1+\sigma^2}\left(1-\frac{r^2}{2b^2}\right)\right]\exp\left(-\frac{r^2}{b^2}\right). \tag{53}$$

In this approximation, behind the FPI the intensity distribution preserves the Gaussian form with a high accuracy. However, in general case the beam pattern is distorted. Note also that the spectral characteristics of the transmission are spatially inhomogeneous and differ for different points of the beam cross section. For example, the wavenumber corresponding to the maximum transmission can be determined from (53) in the form

$$\sigma_{\max}(\mathbf{r}) = 2\alpha_G \left(1 - \frac{r^2}{2b^2}\right). \tag{54}$$

Eqs. (53) and (54) are valid for $r \ll b$. Remarkably, the shift of the transmission maximum in the beam center ($r = 0$) is twice larger than the shift of the integral transmission maximum (41).

**4. It would be useful to confront** the results presented in this paper with the results of other works considering similar problems [4–9]. In [4–6], the transmission band of the interference filters in converging and inclined beams was calculated numerically. Quantitative comparison with their conclusions is difficult because in those works the calculations were performed for high angles of incidence, and the initial data are not always indicated properly; nevertheless, the qualitative agreement is obvious. The analytical calculation of [7] leads to the expression that is a special case of our formula (14); that is why the statement of the authors of [7] that their calculations are valid up to the incidence angles ~30° is, in our opinion, doubtful. In [9], the authors considered the FPI transmittance in the beam that forms the image of a non-coherent source. The APS of such a beam can be determined as a sum of the APSs of the conical beams converging to each point of the source and based on the exit pupil contour of the objective. Substitution of this APS into Eq. (14) will give the expression (1) of [9], which thus appears as a special case of (14).

The spatial distribution of the optical field formed after a Gaussian beam passes the FPI was studied in [8]. Positions of the interference maximum following from Eq. (6) of [8] completely coincide with (54), and the relation (53) looks as a special case of Eq. (6) of [8] that corresponds to condition $\alpha_G \to 0$. Such agreement with the results obtained by completely different method is an additional confirmation of the relevance and expediency of the approach proposed here.

The asymmetry of the transmission contour was not revealed in [8], probably, because the method used by the authors of [8] did not permit an analytical investigation and the situations studied numerically corresponded to small values of $\alpha_G$. Under conditions accepted in [8] $\alpha_G = 0.388$ and $\alpha_G = 0.0388$ while Fig. 1b indicates that the transmission contour asymmetry is only noticeable when $\alpha_G > 1$.

In summary, we emphasize that the results of this study can find possible applications for the design of optical systems employing the FPI in oblique and converging beams as well as in optical metrology. Since the transmission band contour is determined by the quantities $c$ and $\alpha$, i.e. by the set consisting of $q_0$ (31a), (31b), $p$ (17), (18), (29a) and $\gamma$ (26), the band-shape measurements enable finding the third quantity while the two other are known. In this manner, not only the FPI characteristics can be measured but also the parameters of the incident beam APS, especially when employing the conditions of high band-contour sensitivity to these parameters. Other applications can utilize the peculiar features of the transmission contour; for example, the rapid fall-off of the transmittance with growing $\alpha$ at $c > 1$ (see Fig. 2a) can be used for the selection of plane waves with a certain prescribed direction of propagation.